\documentclass[traditabstract]{aa} 

\usepackage{graphicx}
\usepackage{txfonts}
\usepackage{longtable}
\usepackage{natbib}
\begin{document}

   \title{Parent population of flat-spectrum radio-loud \\
	narrow-line Seyfert 1 galaxies}

   \author{M. Berton
	\inst{1}\thanks{marco.berton.1@studenti.unipd.it}
 	  \and
	L. Foschini\inst{2}
	\and
	  S. Ciroi\inst{1}
	   \and
	 V. Cracco\inst{1} 
	\and G. La Mura\inst{1}
	\and M.L. Lister\inst{3} 
	\and \\ S. Mathur \inst{4} \and B.M. Peterson\inst{4} \and J.L. Richards\inst{3} 
	\and P. Rafanelli\inst{1}	
          }

   \institute{$^{1}$ Dipartimento di Fisica e Astronomia "G. Galilei", Universit\`a di Padova, Vicolo dell'Osservatorio 3, 35122, Padova, Italy;\\
 $^{2}$ INAF - Osservatorio Astronomico di Brera, via E. Bianchi 46, 23807 Merate (LC), Italy;\\
	$^{3}$ Department of Physics and Astronomy, Purdue University, 525 Northwestern Avenue, West Lafayette, IN 47907, USA.\\
	$^{4}$ Department of Astronomy and Center for Cosmology and AstroParticle Physics, The Ohio State University, 140 West 18th Avenue, Columbus, OH 43210, USA;\\
             }


\authorrunning{Berton, M. et al.}
\titlerunning{The parent population of F-NLS1s}

\abstract{Flat-spectrum radio-loud narrow-line Seyfert 1 galaxies (NLS1s) are a recently discovered class of $\gamma$-ray emitting Active Galactic Nuclei (AGN), that exhibit some blazar-like properties which are explained with the presence of a relativistic jet viewed at small angles. When blazars are observed at larger angles they appear as radio-galaxies, and we expect to observe an analogue parent population for beamed NLS1s. However, the number of known NLS1s with the jet viewed at large angles is not enough. Therefore, we tried to understand the origin of this deficit. Current hypotheses about the nature of parent sources are steep-spectrum radio-loud NLS1s, radio-quiet NLS1s and disk-hosted radio-galaxies. To test these hypotheses we built three samples of candidate sources plus a control sample, and calculated their black hole mass and Eddington ratio using their optical spectra. We then performed a Kolmogorov-Smirnov statistical test to investigate the compatibility of our different samples with a beamed population. Our results indicate that, when the inclination angle increases, a beamed source appears as a steep-spectrum radio-loud NLS1, or possibly even as a disk-hosted radio-galaxy with low black hole mass and high Eddington ratio. Further investigations, involving larger complete samples and observations at radio frequency, are needed to understand the incidence of disk-hosted radio-galaxies in the parent population, and to assess whether radio-quiet NLS1s can play a role, as well.}

\maketitle

\section{Introduction}
Narrow-line Seyfert 1 galaxies (NLS1s) are a particular subclass of active galactic nuclei (AGN). First classified according to their optical spectral properties by Osterbrock \& Pogge (1985), they later revealed many other interesting aspects at all frequencies. By definition, their full width at half maximum (FWHM) of H$\beta$ is lower than 2000 km s$^{-1}$ \cite{Goodrich89}, the ratio [OIII] $\lambda5007$/H$\beta$ $<$ 3, and their spectra show strong FeII multiplets, signs that the broad-line region (BLR) and the accretion disk are directly visible as in other type 1 AGN. The low FWHM of the permitted lines is usually interpreted as a low orbital velocity around a relatively small central black hole (BH), typically 10$^{6-8}$ M$_\odot$ \cite{Mathur01}. The BH mass is also responsible for a high Eddington ratio $\epsilon$, defined as:
\begin{equation}
\epsilon = \frac{L_{\mathrm{bol}}}{1.3\times10^{38} M_{\mathrm{BH}}/M_\odot} \, ,
\end{equation}
where L$_{bol}$ is the bolometric luminosity. Its value for NLS1s is usually between 0.1 and 1 (Boroson \& Green 1992, Williams et al. 2002, 2004).\par
At radio frequencies the properties of AGN are often described by the radio-loudness parameter $R$, defined as the ratio between the 5 GHz to the optical B-band flux, $F_{5 \, GHz}/F_{B-band}$ \cite{Kellermann89}. The objects with $R$ $<$ 10 are considered radio-quiet, and the majority of NLS1s (93\%, Komossa et al. 2006) belong to this category. However, a few NLS1s are exceptional and show a behavior that is more similar to that of blazars than that of regular Seyfert galaxies, such as a flat radio spectrum and a high brightness temperature (Remillard et al. 1986, Grupe 2000, Zhou \& Wang 2002, Komossa et al. 2006, Whalen et al. 2006, Yuan et al. 2008, from now on Y08). In recent years, following the launch of the \textit{Fermi $\gamma$-ray Space Telescope}, several NLS1s were discovered to be $\gamma$-ray sources (Abdo et al. 2009a,b), confirming the presence of a relativistic beamed jet propagating from their inner core. But, as previously pointed out, the physical characteristics of NLS1s are different from those of BL Lacs and FSRQs. The BH mass is on average two orders of magnitude lower, and the Eddington ratio is higher, usually similar to that of the most powerful quasars. Moreover, the host galaxies of NLS1s are usually of late type, often barred spirals \cite{Crenshaw03}, while blazars are usually hosted in early-type hosts \cite{Sikora07}. \par
The $\gamma$-ray emission in NLS1s, until now, has always been detected in flat-spectrum radio-loud NLS1s (F-NLS1s). The largest sample of these sources known to date is that published by Foschini et al. (2015, from now on F15). Of their 42 objects, 21 have no measured radio spectral index and are assumed to be flat-spectrum sources, while only 7 are actually detected in $\gamma$-rays (F15). From simple geometrical considerations, the number of objects a randomly oriented jet (the parent population) must be $\sim$2$\Gamma^2$ times the number of beamed objects \cite{Urry95}. By taking into account a typical bulk Lorentz factor of 10 (e.g., Abdo et al. 2009b), we expect to find $\sim$8400 parent sources if all of the F15 sources are actually F-NLS1s. Considering instead only those 21 sources with a measured flat spectrum, the parent sources would be $\sim$4200, while in the worst case, if we take into account only the objects with a $\gamma$-ray detection, we expect $\sim$1400 parent sources. However, detailed VLBI studies performed by many authors (Doi et al. 2007, 2011, 2012, Abdo et al. 2009a, Gu \& Chen 2010, Gliozzi 2010, Richards \& Lister 2015) have found only a handful of radio-loud NLS1s (RLNLS1s) with the jet seen at large angles. A clear picture of the problem emerges from studying the complete sample of RLNLS1s reported by Y08. It consists of 23 sources, 12 of which have a flat spectrum and only 5 which have a steep spectrum. The remaining 6 sources have no measured spectral index. Even if we consider these 6 as bona fide steep-spectrum RLNLS1s (S-NLS1s), it is clear from these numbers that something is missing. \par
\begin{figure}[t!]
\centering
\includegraphics[trim=0cm 1.5cm 0cm 0cm, clip=true, width=\hsize]{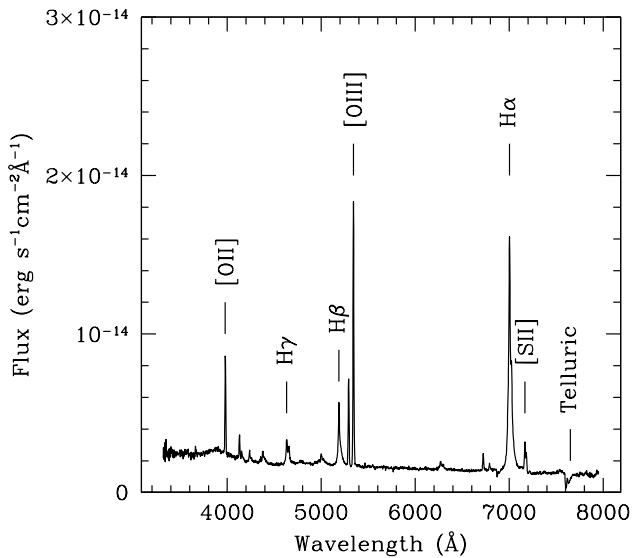}
\caption{Spectrum of RLNLS1 J1302+1624 obtained with the Asiago Astrophysical Observatory 1.22m telescope.}
\label{Mrk335}
\end{figure}
To explain the nature of the parent sources, Foschini (2011, 2012) proposed three options: S-NLS1s, radio-quiet NLS1s (RQNLS1s) and broad-line or narrow-line radio galaxies (BLRGs/NLRGs). The first option, as mentioned before, is numerically inadequate to represent the whole parent population. A second option that might fill the gap is that the parent population is made of RQNLS1s. Many authors have suggested that NLS1s are actually young objects still growing \cite{Mathur01,Grupe00}; if this is true and a jet is present, they might not yet have developed extended radio-lobes and might be also very collimated. Therefore, when observed at large angles, they would become almost invisible for the present-day observatories. Signs of relativistic jets in radio-quiet NLS1 have often been found \cite{Giroletti09,Tarchi11,Doi13,Schonell14}, and a few also have elongated radio-structures in their inner core \cite{Moran00}. A third hypothesis is based on a different assumption about the true nature of NLS1s. Some authors \cite{Decarli08,Risaliti11,Shen14} believe that NLS1s can be due to an orientation effect of a disk-like shaped BLR. When this is observed pole-on, there is no Doppler broadening and the FWHM of the permitted lines is then narrower than in a regular Seyfert 1. In contrast, when observed edge-on, the lines are as broad as usual. In this way, an NLS1 with a beamed jet observed at a different angle would become a regular Seyfert galaxy (type 1 or 2 as always depending on the obscuration) but, because of its radio emission, it would be classified as a broad- or narrow-line radio galaxy. Since NLS1s seem always to be hosted in disk galaxies, the BLRG or NLRG should also be hosted in a disk galaxy (both spiral and lenticular). \par
This work is part of a larger multiwavelength study in preparation that will investigate the physical properties of the parent population of F-NLS1s. The aim of this paper is to search for these sources, focusing on the BH masses and Eddington ratios for three samples of objects corresponding to the previously explained scenarios, and comparing these properties to investigate the relations between the different groups. To do so we will analyze their optical spectra, in particular the H$\beta$ and [OIII] $\lambda$5007 lines. In Sect.~2 is presented the sample selection, in Sect.~3 the data analysis, in Sect.~4 the mass and Eddington ratio calculations, in Sect.~5 the discussion of the results, and in Sect.~6 we briefly summarize. Throughout this work, we adopt a standard $\Lambda$CDM cosmology, with a Hubble constant $H_0 = 70$ km s$^{-1}$ Mpc$^{-1}$, and $\Omega_\Lambda = 0.73$. 
\begin{table*}[t!]
\caption{Steep-spectrum radio-loud NLS1s sample. (1) Short name of the object; (2) alias from NED; (3) right ascension in J2000; (4) declination in J2000; (5) redshift; (6) column density of hydrogen, in units of $10^{21}$ cm$^{-2}$ \cite{Kalberla05}; (7) radio-loudness; (8) source of the optical spectrum: S for SDSS DR9, A for Asiago telescope, T for Telescopio Nazionale Galileo, P for PDF, N for NED.}             
\label{table:1}      
\centering                          
\begin{tabular}{c c c c c c c c}        
\hline\hline                 
Name & Alias NED & R.A. & Dec. & z & nH & RL & Spectrum \\    
\hline                        
\hline
J0146$-$0040 & 2MASX J01464481$-$0040426 & 01h46m44.8s & $-$00d40m43s & 0.083 & 0.291 & 13 & \textbf{S} \\
J0559$-$5026 & PKS 0558$-$504 & 05h59m47.4s & $-$50d26m52s & 0.137 & 0.346 & 26 & P \\
J0806+7248 & RGB J0806+728 & 08h06m38.9s & +72d48m20s & 0.098 & 0.299 & 82 & A \\
J0850+4626 & SDSS J085001.17+462600.5 & 08h50m01.2s & +46d26m01s & 0.524 & 0.267 & 272 & S \\
J0952$-$0136 & Mrk $1239$ & 09h52m19.1s & $-$01d36m43s & 0.020 & 0.341 & 16 & A\\
J1034+3938 & KUG 1031+398 & 10h34m38.6s & +39d38m28s & 0.042 & 0.114 & 33 & S\\
J1200$-$0046 & SDSS J120014.08$-$004638.7 & 12h00m14.1s & $-$00d46m39s & 0.179 & 0.210 & 172 & S \\
J1302+1624 & Mrk $783$ & 13h02m58.8s & +16d24m27s & 0.067 & 0.188 & 23 & A\\
J1305+5116 & SDSS J130522.74+511640.2 & 13h05m22.7s & +51d16m40s & 0.788 & 0.094 & 73 & S \\ 
J1413$-$0312 & NGC 5506 & 14h13m14.9s & $-$03d12m27s & 0.006 & 0.509 & 483 & A \\
J1432+3014 & SDSS J143244.91+301435.3 & 14h32m44.9s & +30d14m35s & 0.355 & 0.120 & 577 & S \\
J1435+3131 & SDSS J143509.49+313147.8 & 14h35m09.5s & +31d31m48s & 0.502 & 0.113 & 6998 & S \\
J1443+4725 & SDSS J144318.56+472556.7 & 14h43m18.5s & +47d25m57s & 0.706 & 0.146 & 1331 & S \\
J1450+5919 & SDSS J145041.93+591936.9 & 14h50m41.9s & +59d19m37s & 0.202 & 0.081 & 30 & S \\
J1703+4540 & SDSS J170330.38+454047.1 & 17h03m30.4s & +45d40m47s & 0.060 & 0.253 & 151 & A\\
J1713+3523 & FBQS J1713+3523 & 17h13m04.5s & +35d23m33s & 0.083 & 0.246 & 73 &S \\
J1722+5654 & SDSS J172206.03+565451.6 & 17h22m06.0s & +56d54m52s & 0.426 & 0.209 & 429 & S \\
J2314+2243 & RX J2314.9+2243 & 23h14m55.7s & +22d43m25s & 0.169 & 0.653 & 17 & A \\
\hline                                   
\end{tabular}
\end{table*}
\begin{table*}[t!]
\caption{Radio-quiet NLS1s sample. Columns as in Table \ref{table:1}.}             
\label{table:2}      
\centering                          
\begin{tabular}{c c c c c c c c}        
\hline\hline                 
Name & Alias NED & R.A. & Dec. & z & nH & RL & Source \\    
\hline                        
\hline
J0044+1921 & RGB J0044+193& 00h44m59.1s & +19d21m41s & 0.181 & 0.316 & 5.6 & A \\
J0632+6340 & UGC $3478$ & 06h32m47.2s & +63d40m25s & 0.013 & 0.676 & $<$10 & T \\
J0752+2617 & RX J0752.7+2617 & 07h52m45.6s & +26d17m36s & 0.082 & 0.340 & 1.6 & S \\
J0754+3920 & B3 0754+394 & 07h58m00.0s & +39d20m29s & 0.096 & 0.512 & 3.0 & A \\
J0913+3658 & RX J0913.2+3658 & 09h13m13.7s & +36d58m17s & 0.107 & 0.147 & 2.9 & S \\
J0925+5217 & Mrk $110$ & 09h25m12.9s & +52d17m11s & 0.035 & 0.131 & 2.0 & A \\
J0926+1244 & Mrk $705$ & 09h26m03.3s & +12d44m04s & 0.029 & 0.357 & 2.4 & A  \\
J0948+5029 & Mrk $124$ & 09h48m42.6s & +50d29m31s & 0.056 & 0.115 & 6.1 & S  \\
J0957+2444 & RX J0957.1+2433 & 09h57m07.2s & +24d33m16s & 0.082 & 0.320 & 1.6 & S \\
J1016+4210 & RX J1016.7+4210 & 10h16m45.1s & +42d10m25s & 0.056 & 0.112 & 0.9 & S \\
J1025+5140 & Mrk $142$ & 10h25m31.3s & +51d40m35s & 0.045 & 0.129 & 0.3 & S\\
J1121+5351 & SBS 1118+541 & 11h21m08.6s & +53d51m21s & 0.103 & 0.095 & 1.9 & S \\
J1203+4431 & NGC $4051$ & 12h03m09.6s & +44d31m53s & 0.002 & 0.114 & 3.1 & A \\
J1209+3217 & RX J1209.7+3217 & 12h09m45.2s & +32d17m01s & 0.144 & 0.134 & 4.2 & S \\
J1215+5242 & SBS 1213+549A & 12h15m49.4s & +54d42m24s & 0.150 & 0.155 & 4.0 & S\\
J1218+2948 & Mrk $766$ & 12h18m26.5s & +29d48m46s & 0.013 & 0.188 & 7.6 & A \\
J1242+3317 & WAS $61$ & 12h42m10.6s & +33d17m03s & 0.044 & 0.143 & 4.1 & S \\
J1246+0222 & PG 1244+026 & 12h46m35.2s & +02d22m09s & 0.048 & 0.168 & 1.3 & S \\
J1337+2423 & IRAS 13349+2438 & 13h37m18.7s & +24d23m03s & 0.108 & 0.100 & 4.0 & A \\
J1355+5612 & SBS 1353+564 & 13h55m16.5s & +56d12m45s & 0.122 & 0.100 & 7.9 & S\\
J1402+2159 & RX J1402.5+2159 & 14h02m34.4s & +21d59m52s & 0.066 & 0.195 & 1.7 & A \\
J1536+5433 & Mrk $486$ & 15h36m38.3s & +54d33m33s & 0.039 & 0.144 & 0.5 & A \\
J1537+4942 & SBS 1536+498 & 15h37m32.6s & +49d42m48s & 0.280 & 0.169 & 9.6 & S \\
J1555+1911 & Mrk $291$ & 15h55m07.9s & +19d11m33s & 0.035 & 0.285 & 1.9 & S \\
J1559+3501 & Mrk $493$ & 15h59m09.6s & +35d01m47s & 0.031 & 0.213 & 3.8 & A \\
\hline
\end{tabular}
\end{table*}
\begin{table*}[t!]
\caption{Radio galaxies with a disk host sample. Columns as in Table \ref{table:1}. Sources from the sample of Inskip et al. (2010) are marked with an asterisk.}             
\label{table:3}      
\centering                          
\begin{tabular}{c c c c c c c c}        
\hline\hline                 
Name & Alias NED & R.A. & Dec. & z & nH & RL & Spectrum \\    
\hline                        
\hline
J0010+1058 & Mrk $1501$ & 00h10m31.0s & +10d58m30s & 0.089 & 0.574 & 315 & A \\
J0150$-$0725 & F 01475$-$0740 & 01h50m02.7s & $-$07d25m48s & 0.018 & 0.203 & 1121 & A \\
J0316+4119 & IC $310$ & 03h16m43.0s & +41d19m30s & 0.019 & 1.240 & 298 & S \\ 
J0407+0342 & 3C $105$* & 04h07m16.5s & +03d42m26s & 0.089 & 1.090 & 115258 & N \\
J0433+0521 & 3C $120$* & 04h33m11.1s & +05d21m16s & 0.033 & 1.020 & 2181 & A \\
J0552$-$0727 & NGC $2110$ & 05h52m11.4s & $-$07d27m22s & 0.008 & 1.620 & 5372 & A \\
J0725+2957 & B2 0722+30 & 07h25m37.3s & +29d57m15s & 0.019 & 0.590 & 242 & A \\
J1140+1743 & NGC $3801$ & 11h40m16.9s & +17d43m41s & 0.011 & 0.209 & 4101 & S \\
J1252+5634 & 3C $277.1$ & 12h52m26.3s & +56d34m20s & 0.320 & 0.080 & 12646 & S \\
J1312+3515 & PG 1309+355 & 13h12m17.8s & +35d15m21s & 0.183 & 0.100 & 40 & S \\
J1324+3622 & NGC $5141$ & 13h24m51.4s & +36d22m43s & 0.017 & 0.101 & 564 & S \\
J1352+3126 & UGC $8782$ & 13h52m17.8s & +31d26m46s & 0.045 & 0.126 & 21185 & S\\
J1409$-$0302 & SDSS J140948.85$-$030232.5 & 14h09m48.8s & $-$03d02m33s & 0.137 & 0.463 & 60 &S\\
J1449+6316 & 3C $305$ & 14h49m21.6s & +63d16m14s & 0.042 & 0.138 & 2994 & S \\
J1550+1120 & SDSS J155043.59+112047.4 & 15h50m43.6s & +11d20m47s & 0.436 & 0.351 & 5397 & S\\
J1704+6044 & 3C $351$ & 17h04m41.4s & +60d44m31s & 0.372 & 0.169 & 2584 & A \\
\hline
\end{tabular}
\end{table*}
\begin{table*}[t!]
\caption{Control sample: radio galaxies with an elliptical host. Columns as in Table \ref{table:1}.}             
\label{table:4}      
\centering                          
\begin{tabular}{c c c c c c c c}        
\hline\hline                 
Name & Alias NED & R.A. & Dec. & z & nH & RL & Spectrum \\    
\hline                        
\hline
J0037$-$0109 & 3C $15$ & 00h37m04.1s & $-$01d09m08s & 0.073 & 0.223 & 1005 & N \\
J0038$-$0207 & 3C $17$ & 00h38m20.5s & $-$02d07m41s & 0.220 & 0.285 & 63550 &N \\
J0040+1003 & 3C $18$ & 00h40m50.5s & +10d03m23s & 0.188 & 0.556 & 2376 & N \\
J0057$-$0123 & 3C $29$ & 00h57m34.9s & $-$01d23m28s & 0.045 & 0.328 & 1237 & N \\
J0327+0233 & 3C $88$ & 03h27m54.2s & +02d33m42s & 0.030 & 0.809 & 14646 & N \\
J0808$-$1027 & 3C $195$ & 08h08m53.6s & $-$10d27m40s & 0.109 & 0.760 & 11905 & N \\
J0947+0725 & 3C $227$ & 09h47m45.1s & +07d25m20s & 0.086 & 0.204 & 9909 &A \\
J1602+0157 & 3C $327$ & 16h02m27.4s & +01d57m56s & 0.105 & 0.576 & 16692 &A \\
J1952+0230 & 3C $403$ & 19h52m15.8s & +02d30m24s & 0.059 & 1.130 & 16963 & N \\
J2223$-$0206 & 3C $445$ & 22h23m49.5s & $-$02d06m13s & 0.056 & 0.484 & 3407 &N \\
J2316+0405 & 3C $459$ & 23h16m35.2s & +04d05m18s & 0.220 & 0.550 & 11382 &N \\
\hline
\end{tabular}
\end{table*}
\section{Sample selection}
To check the three different hypotheses about the parent population of F-NLS1s, we created three samples and one control sample. To improve the otherwise small number statistics, we chose to add every source we found in the literature, at the expense of having incomplete samples. Since our aim is to use the H$\beta$ and [OIII] $\lambda$5007 \AA{} lines to estimate the BH mass and the Eddington ratio, we limited our study to objects with an optical spectrum that we could analyze. \par
We calculated the radio-loudness for each source. For 22 of them the only radio flux available was at 1.4 GHz, therefore we extrapolated their 5 GHz flux in approximation of a typical pure synchrotron steep spectrum with spectral index $\alpha = 0.7$ ($F_\nu \propto \nu^{-\alpha}$). The B-band magnitude, when possible, was derived directly from the optical spectrum by convolving it with a B-filter template and then calculating the integrated flux. Otherwise, we retrieved the B magnitude from NED\footnote{http://ned.ipac.caltech.edu} and SIMBAD\footnote{http://simbad.u-strasbg.fr/simbad/} archives.\par
\textbf{RLNLS1s:} The first group consists of 18 NLS1s with a steep radio spectral index $\alpha > 0.5$ and a radio-loudness $R > 10$. The sample was selected by using all the sources found in previous surveys \cite{Zhou02,Wadadekar04,Komossa06,Whalen06,Yuan08} and from individual studies \cite{Gliozzi10,Tarchi11,Caccianiga14}. All the sources are classified as NLS1s according to their FWHM(H$\beta$) $<$ 2000 km s$^{-1}$, their ratio [OIII]/H$\beta$ $<$ 3, and the presence of FeII multiplets. The only outlier, J1413$-$0312, is described in Appendix A.\par
\textbf{RQNLS1s:} The second group, 25 radio-quiet - but not radio-silent - NLS1s with $R < 10$, includes all the sources detected at 1.4 GHz by the FIRST survey, as reported in Wadadekar (2004). As for RLNLS1s, all the sources are classified as NLS1s according to their optical spectrum. \par
\textbf{BLRG/NLRG:} The third group, 16 disk-hosted radio galaxies (RGs), was selected from data available in the literature. To confirm the host galaxy nature, we cross-checked the classifications found in the HyperLeda database\footnote{http://leda.univ-lyon1.fr} \cite{Paturel03} with what we found in literature. We kept only sources with a confirmed classification. For a description on specific objects, see Appendix A. \par
As control sample we chose to use the flux-limited sample of 2 Jy RGs defined by Inskip et al. (2010). Their 43 sources have a flux density F$_{2.7 GHz} >$ 2 Jy and a declination $\delta < 10^\circ$. 12\% of them are found to be hosted in disk galaxy in that same work, and therefore we included them in the third group. The others are objects hosted by elliptical galaxies. We obtained an optical spectrum for 11 of them, and used these as control sample (see Sect.3). The samples are listed in Tables \ref{table:1}$-$\ref{table:4}. Finally we used the sample of F-NLS1s shown in Table 1 of F15 to compare the physical properties of our candidate parent sources with those of the beamed population.
\section{Data analysis}
\begin{table}
\caption{Observational details for Asiago optical spectra. (1) Object name; (2) exposure time in seconds; (3) rest frame spectral coverage (\AA{}).}             
\label{table:5}      
\centering                          
\begin{tabular}{c c c}        
\hline\hline                 
Name & Exposure time (s) & Spectral coverage (\AA{}) \\    
\hline                       
\hline
J0010$+$1058 & 4800 & 2960$-$7280 \\
J0044$+$1921 & 9600 & 2900$-$6700 \\
J0150$-$0725 & 2400 & 3590$-$7760 \\
J0433$+$0521 & 3600 & 3120$-$7680 \\
J0552$-$0727 & 2400 & 3160$-$7840 \\
J0725$+$2957 & 4800 & 3360$-$7750 \\
J0754$+$3920 & 4800 & 2900$-$7200 \\
J0806$+$7248 & 4800 & 2940$-$7220 \\
J0925$+$5217 & 2400 & 3220$-$7780 \\
J0926+1244 & 4800 & 3210$-$7680 \\
J0947+0725 & 4800 & 3150$-$7270 \\
J0952$-$0136 & 4800 & 3130$-$7750 \\
J1203+4431 & 2400 & 3230$-$7940 \\
J1218+2948 & 13200 & 3420$-$7860 \\
J1302+1624 & 3600 & 3050$-$7450 \\
J1337+2423 & 4800 & 3120$-$7180 \\
J1402+2159 & 4800 & 3030$-$7460 \\
J1413$-$0312 & 2400 & 3210$-$7890 \\
J1536+5433 & 2400 & 3210$-$7760 \\
J1559+3501 & 3600 & 3130$-$7720 \\
J1602+0157 & 4800 & 2930$-$7200 \\
J1703+4540 & 1200 & 3220$-$7450 \\
J1704+6044 & 4800 & 2440$-$5800 \\
J2314+2243 & 7200 & 2820$-$6760 \\
J2345$-$0449 & 4800 & 3170$-$7340 \\
\hline
\end{tabular}
\end{table}
The primary sources for the optical spectra were the Sloan Digital Sky Survey (SDSS) Data Release 9 and the NED archive. For all sources with a declination higher than -15$^\circ$, an apparent magnitude lower than 18 and no published spectrum, we obtained a spectrum using the 1.22m telescope of the Asiago Astrophysical Observatory (Italy). In one case, J0559$-$5026, we converted the optical spectrum from Remillard et al. (1986) into an analyzable FITS format using the \texttt{digitizer} software\footnote{http://digitizer.sourceforge.net/}. The sources of the spectra are reported in Tables \ref{table:1}$-$\ref{table:4}. \par
The subsequent data reduction was performed using the standard tasks of \texttt{IRAF v.2.14.1}. We collected the spectra at the Asiago 1.22m telescope between April 2013 and September 2014, using the Boller \& Chivens spectrograph with a 300 mm$^{-1}$ grating. The instrumental resolution was R $\sim$ 700. The slit had an aperture of 4.25'' on the sky plane, a good compromise for obtaining the nuclear spectrum for nearby objects and the whole galaxy spectrum for high-redshift sources. The exposure time and the rest frame spectral coverage for each object is reported in Table \ref{table:5}; we split observations into exposures of 1200 s or 1800 s each, to avoid a strong contamination by cosmic rays and light pollution. In the pre-reduction we used overscan instead of bias, and NeHgAr or FeAr lamps were used for the wavelength calibration. After the flux calibration and the sky subtraction, we extracted monodimensional spectra for each object, and later combined them (e.g., see Fig. \ref{Mrk335}). In one case, J0632+6340, we obtained the optical spectrum in October 2005 using the 3.58m Telescopio Nazionale Galileo (TNG), with the DOLORES camera (device optimized for the low resolution). We used the MR-B Grm2 grism with a 1.1'' slit with a resolution R $\sim$ 2100. The exposure time was 6100 s, and a He lamp was used to perform the wavelength calibration \cite{Berton10}. \par
All the flux calibrated spectra were first corrected for Galactic absorption using the nH values reported in Kalberla et al. (2005), and were then corrected for redshift. The host galaxy contribution is negligible in many objects; in fact, for sources with redshift $z > 0.1$, the host component is lower than 10\% of the whole spectrum \cite{Letawe07}. In closer objects we examined the spectra for signs of stellar absorption. In most cases the AGN continuum and lines were still much stronger than those coming from the host, particularly for type 1 objects, and the host subtraction had no influence on the line profiles, so we continued the analysis without subtracting its contribution, as in F15. In type 2 or intermediate objects the absorptions were often clearly visible, and we subtracted an adequate host galaxy template \cite{Kinney96}, according to the morphological classification and spectral shape of each object. \par
\begin{figure}[t!]
\centering
\includegraphics[trim=0cm 2.0cm 0cm 0cm, clip=true, width=\hsize]{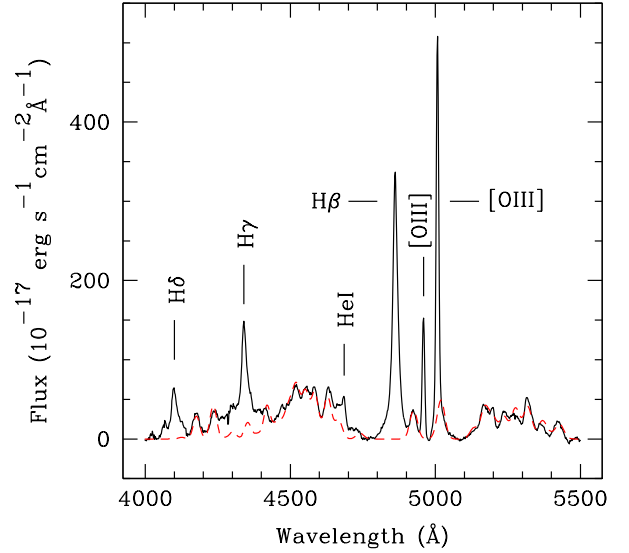}
\caption{Spectrum of J0632+6340 (solid black line) obtained with the TNG, with a FeII template obtained from the online software (dashed red line).}
\label{fig:fe}
\end{figure}

We focused our analysis on the H$\beta$ region, between 4000 and 5500 \AA{}: when the FeII multiplets were present, we subtracted them using the online software\footnote{http://servo.aob.rs/FeII\_AGN/} developed by Kova{\v c}evi{\'c} et al. (2010) and Shapovalova et al. (2012). This software provides a best-fit model that reproduces the iron multiplets in the H$\beta$ region for each object as function of gas temperature, Doppler broadening, and shift of the FeII lines. An example of a template is shown in Fig.\ref{fig:fe}. We then proceeded in two different ways for type 1 AGN and intermediate or type 2 objects. 
\begin{itemize}
\item{\textbf{Sy1:}} In this first case, we decomposed the H$\beta$ line into three Gaussian components, one to reproduce the narrow component with the task \texttt{ngaussfit} of \texttt{IRAF}, and two more for the broad component. The center of the narrow component was always free to vary, and as suggested in V{\'e}ron-Cetty et al. (2001), we fixed its flux to 1/10 of that of [OIII] $\lambda$5007, the mean value for Seyfert galaxies, and its FWHM to that of [OIII]. Nevertheless, these parameters did not always provide a satisfactory result in fitting the line profile: the gas in which the [OIII] originates is often turbulent, as indicated by the recurring presence of blue wings in the line profile, and the line global width can lead to an overestimate of the narrow component. For this reason we used the core component of the [OIII] line as a reference, or we let the permitted narrow component width vary freely below the core width. Moreover, when the fit result was clearly incorrect, we also let the narrow component flux vary freely. In some cases we fitted the line with only two Gaussians, one broad and one narrow, because of an irregular H$\beta$ profile. Once we obtained the best fit, we subtracted the narrow component and measured the line dispersion $\sigma$, defined as the second-order momentum of the broad component. The use of $\sigma$ instead of the FWHM gives better results for low-contrast lines, and also a lower uncertainty \cite{Peterson11}. We did not correct for the instrumental resolution, because even in the narrowest H$\beta$ the effect of this correction is negligible. 
\item{\textbf{Sy2/Sy-intermediate:}} In the other sources we followed the same initial steps, but because of the obscuration due to the molecular torus, we could not use the H$\beta$ broad line. Another way to determine the BH mass is to exploit its relation with the stellar velocity dispersion. However, in our objects the stellar absorption lines are almost invisible in the nuclear spectra, so we could not directly derive $\sigma_*$, and we used the forbidden lines instead. Low-ionization lines, such as [NII], are probably more suitable for this purpose, but they are not always present in our spectra because of the redshift, so we focused on the [OIII] $\lambda\lambda$4959, 5007 doublet. We decomposed them using two Gaussians each, the first one representing a core component, and the second one representing the secondary (often blueshifted) component. To reduce the number of free parameters, we fixed both FWHMs to be the same in the two lines, and the flux of the $\lambda$4959 line to be one-third of the $\lambda$5007 line, as predicted by the theory. In objects where the $\lambda$4959 line was dominated by noise we did not use any constraints and focused on obtaining a good fit to the $\lambda$5007 line alone. Finally, in those cases where the FWHM of the core component was above the instrumental resolution limit, we measured it and corrected it for the instrumental resolution. 
\end{itemize}

\section{Mass and accretion rate}
For all the NLS1s and the type 1 objects, after obtaining the value of $\sigma$ for the H$\beta$ broad component, we calculated the black hole mass under the hypothesis of a virialized system according to
\begin{equation}
M_{BH} = f\left(\frac{R_{\mathrm{BLR}} \sigma^2_{\mathrm{H}\beta}}{G}\right) \, ,
\end{equation}
where $R_{\mathrm{BLR}}$ is the radius of the BLR, $G$ is the gravitational constant, and $f$ is the scaling factor \cite{Peterson04} which, as suggested in Collin et al. (2006), we assumed to be 3.85. This value, as pointed out in their work, is not dependent on the inclination of the BLR, so it can be used in all our NLS1s samples. To find the BLR size, we used the relation developed by Greene et al. (2010) that links it to the H$\beta$ luminosity, 
\begin{equation}
\log\left(\frac{R_{\mathrm{BLR}}}{10 \textrm{ l.d.}}\right) = 0.85 + 0.53\log\left(\frac{L(\textrm{H}\beta)}{10^{43} \textrm{erg s}^{-1}}\right) \, .
\end{equation}
\begin{figure}[t!]
\centering
\includegraphics[trim=0cm 1.5cm 0cm 0cm, clip=true,width=\hsize]{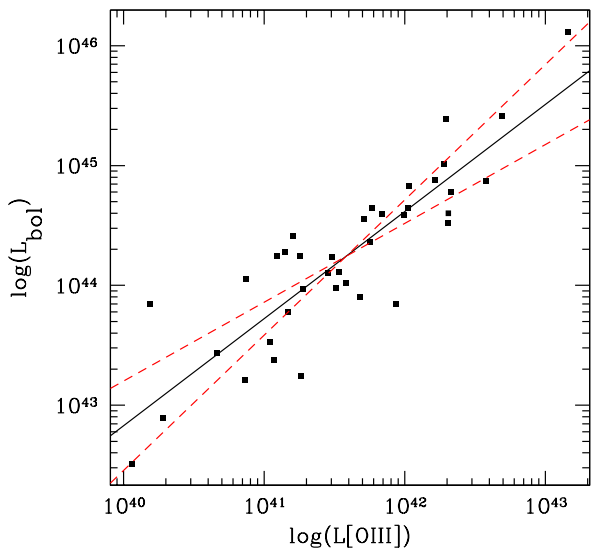}
\caption{Relation between the [OIII] $\lambda$5007\AA{} luminosity and the bolometric luminosity. The black solid line is the best fit, the red dashed lines are the highest and lowest slope lines.}
\label{o3bol}
\end{figure}
By using an emission line to determine the BLR radius, we avoid the possible jet contamination that can affect the continuum luminosity at 5100 \AA{} that is often used in many other studies. The BLR size also provides a way to estimate the disk luminosity: if we assume a photoionization regime, $R_{\mathrm{BLR}} \propto \sqrt{L_{\mathrm{disk}}}$ \cite{Koratkar91,Ghisellini09}. Under the reasonable hypothesis that the bolometric luminosity is comparable with the disk luminosity, we can estimate the Eddington ratio. \par
This second method, as already mentioned, is based on the [OIII] $\lambda$5007\AA{} line. As shown by Nelson \& Whittle (1996) there is a relationship between the [OIII] line width and the stellar velocity dispersion $\sigma_*$ of the galaxy bulge. This relation was also investigated in the work by Greene \& Ho (2005), and they found that the estimates improve when the core component of [OIII] is used instead of the whole FWHM. When both components had a FWHM higher than the instrumental resolution, we focused solely on the core component after decomposing the [OIII] lines. In contrast, when one of the components was unresolved, we measured the FWHM of the entire line. These cases correspond to
\begin{equation}
\label{eqn:1}
\sigma_* = \frac{\textrm{FWHM}^c_{\mathrm{[OIII]}}}{2.35} \; \textrm{and} 
\end{equation}
\begin{equation}
\label{eqn:2}
\sigma_* = \frac{\textrm{FWHM}_{\mathrm{[OIII]}}}{1.34\times2.35} \, .
\end{equation}
As is widely known, $\sigma_*$ is correlated with the black hole mass in the $M_{BH}-\sigma_*$ relation \cite{Ferrarese00}. To estimate the masses, we used the revised relation found by Ho \& Kim (2014):
\begin{equation}
\label{bhsigma}
\log\left(\frac{M_{\mathrm{BH}}}{M_\odot}\right) = 8.49 + 4.38\log\left(\frac{\sigma_*}{200 \, \textrm{km s}^{-1}}\right) \, .
\end{equation}
The best way to obtain the bolometric luminosity for obscured sources is still debated; although it is commonly accepted that there is a relation between the [OIII] line luminosity and the bolometric luminosity \cite[i.e.][]{Heckman04,Wang07,Lamastra09,Risaliti11}, quantitatively it is still uncertain. We therefore decided to calculate a new normalization of the relation using our sample of type 1 objects. As previously explained, we derived the bolometric luminosity of NLS1s from the H$\beta$ luminosity, and we measured the [OIII] luminosity separately. As shown in Fig.\ref{o3bol}, the correlation is evident. To find the best-fit line we used the least-squares method, which led to the following relation: 
\begin{equation}
\label{eqn:o3bol}
\log\left(\frac{L_{\mathrm{bol}}}{\textrm{erg s}^{-1}}\right) = (7.54\pm9.07) + (0.88\pm0.22)\log\left(\frac{L_{\mathrm{[OIII]}}}{\textrm{erg s}^{-1}}\right) \, .
\end{equation}
The dispersion in this relation is 0.2 dex. Under the assumption that the unified model is valid \cite{Antonucci93}, this relation can be used for type 2 or intermediate AGN, even if it was obtained from type 1 objects. We did not correct our data for intrinsic absorption of the galaxies, therefore the bolometric luminosity and the Eddington ratios might be underestimated and should be considered as lower limits. \par
\section{Discussion}
\begin{figure}[t!]
\flushleft
\includegraphics[trim=0cm 1.5cm 0cm 0cm, clip=true,width=\hsize]{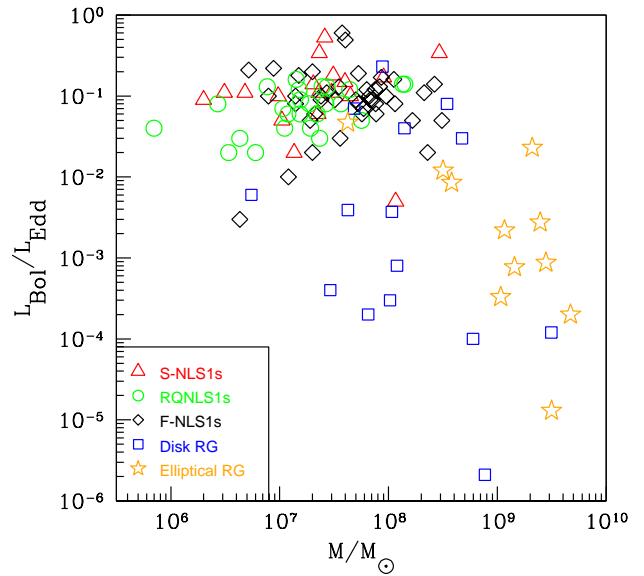}
\caption{BH mass vs Eddington ratio. Red triangles are S-NLS1s, green circles are RQNLS1s, blue empty squares are disk-hosted BLRGs and NLRGs, and orange stars are elliptical RGs. In black we plot the sample of F-NLS1s from F15.}
\label{massacc}
\end{figure}
\begin{figure}[t!]
\centering
\includegraphics[trim=0cm 1.5cm 0cm 0cm, clip=true,width=\hsize]{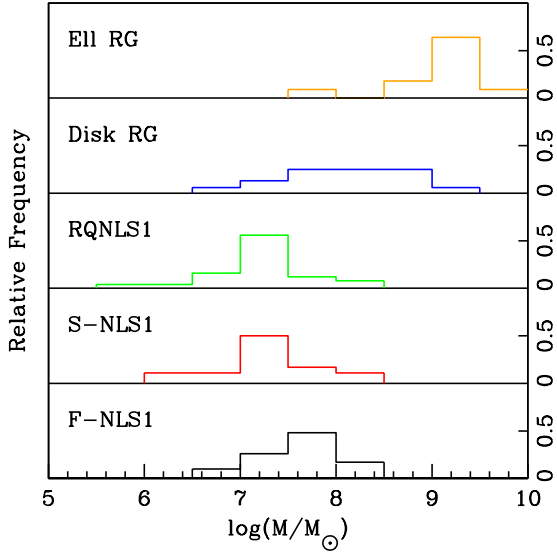}
\caption{Mass distribution of the samples. From bottom to top: (1) In black, flat-spectrum radio-loud NLS1s from F15; (2) in red, steep-spectrum radio-loud NLS1s; (3) in green, radio-quiet NLS1s; (4) in blue, disk-hosted BLRGs and NLRGs; (5) in orange, elliptical RGs. }
\label{masse}
\end{figure}
\begin{figure}
\centering
\includegraphics[trim=0cm 1.5cm 0cm 0cm, clip=true,width=\hsize]{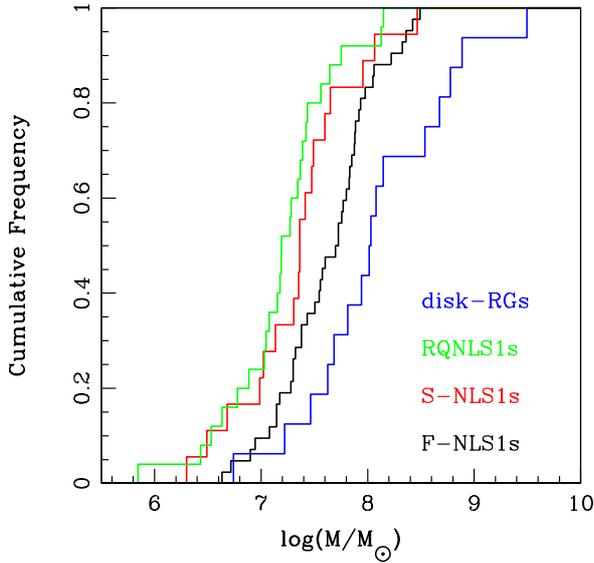}
\caption{Cumulative distributions of the samples. (1) In black, flat-spectrum radio-loud NLS1s from F15; (2) In red, steep-spectrum radio-loud NLS1s; (3) in green, radio-quiet NLS1s; (4) in blue, disk-hosted BLRGs and NLRGs.}
\label{cumulativa}
\end{figure}
\begin{table}
\label{table:mass1}
\caption{Mass and accretion luminosity estimated for NLS1s. Columns: (1) Name of the source; (2) logarithm of the H$\beta$ luminosity; (3) logarithm of the [OIII] $\lambda$5007 \AA{} luminosity; (4) logarithm of the black hole mass; (5) logarithm of the bolometric luminosity; (6) logarithm of the Eddington ratio.} 
\centering
\begin{tabular}{c c c c c c c}
\hline\hline
Name & $\log{\textrm{L}_{H\beta}}$ & $\log{\textrm{L}_{[OIII]}}$ & $\log{\textrm{M}_{BH}}$ & $\log{\textrm{L}_{bol}}$ & $\log\epsilon$ \\
\hline
\textbf{RLNLS1s} & {} & {} & {} & {} \\
\hline
J0146$-$0040 & 41.79 & $-$ & 7.35 & 44.25 & -1.22 \\
J0559$-$5026 & 42.74 & $-$ & 7.41 & 45.25 & -0.28 \\
J0806$+$7248 & 41.41 & $-$ & 6.68 & 43.84 & -0.96 \\
J0850$+$4626 & 42.30 & $-$ & 7.65 & 44.78 & -1.00 \\
J0952$-$0136 & 41.41 & $-$ & 7.02 & 43.84 & -1.30 \\
J1034$+$3938 & 40.95 & $-$ & 6.30 & 43.38 & -1.05 \\
J1200$-$0046 & 42.09 & $-$ & 7.31 & 44.56 & -0.85 \\
J1302$+$1624 & 42.05 & $-$ & 7.36 & 44.52 & -0.96 \\
J1305$+$5116 & 43.56 & $-$ & 8.47 & 46.12 & -0.47 \\
J1413$-$0312 & $-$ & 40.62 & 8.07 & 43.30 & -2.91 \\
J1432$+$3014 & 42.38 & $-$ & 7.49 & 44.87 & -0.74 \\
J1435$+$3131 & 42.16 & $-$ & 7.48 & 44.64 & -0.96 \\
J1443$+$4725 & 42.52 & $-$ & 7.36 & 45.01 & -0.47 \\
J1450$+$5919 & 41.67 & $-$ & 6.99 & 44.11 & -1.00 \\
J1703$+$4540 & 40.30 & $-$ & 6.49 & 43.68 & -0.96 \\
J1713$+$3523 & 41.11 & $-$ & 7.13 & 43.53 & -1.70 \\
J1722$+$5654 & 42.39 & $-$ & 7.60 & 44.88 & -0.82 \\
J2314$+$2243 & 42.79 & $-$ & 7.95 & 45.30 & -0.77 \\
\hline
\textbf{RQNLS1s} & {} & {} & {} & {} & {} \\
\hline
J0044$+$1921 & 42.01 & $-$ & 7.15 & 44.48 & -0.80 \\
J0632$+$6340 & 40.48 & $-$ & 6.53 & 42.89 & -1.70 \\
J0752$+$2617 & 41.83 & $-$ & 7.27 & 44.28 & -1.10 \\
J0754$+$3920 & 42.89 & $-$ & 8.15 & 45.41 & -0.85 \\
J0913$+$3658 & 41.53 & $-$ & 7.08 & 43.98 & -1.22 \\
J0925$+$5217 & 42.11 & $-$ & 7.56 & 44.59 & -1.10 \\
J0926$+$1244 & 41.53 & $-$ & 7.28 & 43.97 & -1.40 \\
J0948$+$5029 & 41.58 & $-$ & 7.03 & 44.02 & -1.15 \\
J0957$+$2433 & 41.65 & $-$ & 6.89 & 44.10 & -0.89 \\
J1016$+$4210 & 41.79 & $-$ & 7.34 & 44.25 & -1.22 \\
J1025$+$5140 & 41.61 & $-$ & 7.19 & 44.05 & -1.22 \\
J1121$+$5351 & 42.34 & $-$ & 7.64 & 44.83 & -0.92 \\
J1203$+$4431 & 40.15 & $-$ & 5.85 & 42.51 & -1.40 \\
J1209$+$3217 & 42.17 & $-$ & 7.44 & 44.65 & -0.89 \\
J1215$+$5242 & 42.12 & $-$ & 7.75 & 44.59 & -1.30 \\
J1218$+$2948 & 40.85 & $-$ & 6.78 & 43.24 & -1.70 \\
J1242$+$3317 & 41.46 & $-$ & 7.37 & 43.90 & -1.52 \\
J1246$+$0222 & 41.34 & $-$ & 7.05 & 43.90 & -1.40 \\
J1337$+$2423 & 42.87 & $-$ & 8.13 & 45.46 & -0.85 \\
J1355$+$5612 & 42.13 & $-$ & 7.39 & 44.60 & -0.89 \\
J1402$+$2159 & 41.78 & $-$ & 7.19 & 44.23 & -1.05 \\
J1536$+$5433 & 41.95 & $-$ & 7.42 & 44.42 & -1.10 \\
J1537$+$4942 & 42.00 & $-$ & 7.18 & 44.36 & -0.92 \\
J1555$+$1911 & 40.78 & $-$ & 6.63 & 43.20 & -1.52 \\
J1559$+$3501 & 41.04 & $-$ & 6.43 & 43.43 & -1.10 \\
\hline
\end{tabular}
\end{table}
Our results are displayed in Tables 6 and \ref{table:mass2}. We compared our findings with the sample of F-NLS1s studied in F15, and all of them are shown together in Figs. \ref{massacc}, \ref{masse} and \ref{cumulativa}. We expect that the flat- and steep-spectrum populations have a similar mass distributions, and they both should be different from that of the control sample of elliptical radio-galaxies. \par
The mass and accretion luminosity values for both NLS1s samples roughly agree with previous results found in the literature \cite[see][]{Jarvela14}. The Eddington ratio $\epsilon$ is quite different only in J1413$-$0312, as discussed in Appendix A. The two samples have similar distributions. The average mass value for RQ sources is 2.8$\times$10$^7$ M$_\odot$, and a median of 1.6$\times$10$^7$ M$_\odot$, while S-NLS1s have an average mass of 4.5$\times$10$^7$ M$_\odot$ and a median of 2.3$\times$10$^7$ M$_\odot$. For both samples the dispersion is 0.8 dex. These two results are quite similar to the average value of 6.2$\times$10$^7$ M$_\odot$ found for F-NLS1s (F15). The difference between the average and median of the two samples is due to the presence of a few high-mass objects in the distributions. \par
The samples of radio-galaxies have different mass distributions that strongly depend on whether the host galaxy is a disk or an elliptical, and this is particularly evident in the histogram of Fig.5. This can be understood in the context of the M$_{BH}-\sigma_*$ relation. The stellar velocity dispersion in disk-galaxy bulges is systematically lower than that in elliptical galaxies, and inevitably leads to a lower BH mass. Disk RGs have average and median masses of 3.8$\times$10$^8$ and 1.1$\times$10$^8$ M$_\odot$, with a dispersion of 0.8 dex. For elliptical RGs, instead, average and median masses are 1.8$\times$10$^9$ and 1.4$\times$10$^9$ M$_\odot$, with a dispersion of 1.2 dex. These values, as expected, are an order of magnitude higher than the others. \par
The Eddington ratio is on average lower in RGs than in NLS1s. This may be due both to a real physical effect and to an underestimate of the bolometric luminosity (see Sect. 4). Despite this, there are a few disk-hosted RGs for which $\epsilon$ is similar to that of NLS1s. A few ellipticals also have a relatively high $\epsilon$, in analogy with the high-mass/high-Eddington ratio typical of FSRQs. The remaining elliptical galaxies are instead located at low $\epsilon$, in a similar way to the BL Lacs \cite{Ghisellini10}.
\begin{table}[t!]
\caption{Mass and accretion luminosity estimated for radio-galaxies. Columns as in table 6.} 
\label{table:mass2}
\begin{tabular}{c c c c c c}
\hline\hline
Name & $\log{\textrm{L}_{H\beta}}$ & $\log{\textrm{L}_{[OIII]}}$ & $\log{\textrm{M}_{BH}}$ & $\log{\textrm{L}_{bol}}$ & $\log\epsilon$ \\
\hline
\textbf{Disk RGs} & {} & {} & {} & {} & {} \\
\hline
J0010$+$1058 & 42.42 & $-$ & 8.15 & 44.91 & -1.40 \\
J0150$-$0725 & $-$ & 40.40 & 8.08 & 43.10 & -3.10 \\
J0316$+$4119 & $-$ & 39.28 & 7.47 & 42.15 & -3.40 \\
J0407$+$0342 & $-$ & 41.09 & 9.50 & 43.70 & -3.92 \\
J0433$+$0521 & 42.19 & $-$ & 7.68 & 44.66 & -1.15 \\
J0552$-$0727 & $-$ & 40.65 & 7.63 & 43.33 & -2.41 \\
J0725$+$2957 & $-$ & 39.87 & 6.74 & 42.63 & -2.22 \\
J1140$+$1743 & $-$ & 38.48 & 8.89 & 41.30 & -5.68 \\
J1252$+$5634 & $-$ & 42.94 & 8.68 & 45.32 & -1.52 \\
J1312$+$3515 & 43.01 & $-$ & 8.54 & 45.54 & -1.10 \\
J1324$+$3622 & $-$ & 39.54 & 7.81 & 42.32 & -3.70 \\
J1352$+$3126 & $-$ & 39.89 & 8.01 & 42.64 & -3.52 \\
J1409$-$0302 & $-$ & 40.31 & 8.78 & 43.01 & -4.00 \\
J1449$+$6316 & $-$ & 41.10 & 8.03 & 43.72 & -2.43 \\
J1550$+$1120 & $-$ & 43.04 & 7.22 & 45.42 & 0.08 \\
J1704$+$6044 & $-$ & 43.04 & 7.94 & 45.41 & -0.64 \\
\hline
\textbf{Elliptical RGs} & {} & {} & {} & {} \\
\hline
J0037$-$0109 & $-$ & 41.54 & 9.67 & 44.09 & -3.70 \\
J0038$-$0207 & $-$ & 39.67 & 9.45 & 44.50 & -3.06 \\
J0040$+$1003 & $-$ & 43.47 & 9.32 & 45.80 & -1.64 \\
J0057$-$0123 & $-$ & 41.05 & 9.03 & 43.67 & -3.48 \\
J0327$+$0233 & $-$ & 39.97 & 9.50 & 42.71 & -4.89 \\
J0808$-$1027 & $-$ & 42.14 & 8.58 & 44.63 & -2.07 \\
J0947$+$0725 & $-$ & 41.90 & 7.63 & 44.41 & -1.33 \\
J1602$+$0157 & $-$ & 42.22 & 8.50 & 44.69 & -1.92 \\
J1952$+$0230 & $-$ & 41.61 & 9.16 & 44.16 & -3.11 \\
J2223$-$0206 & $-$ & 42.51 & 9.39 & 44.95 & -2.56 \\
J2316$+$0405 & $-$ & 42.02 & 9.07 & 44.52 & -2.66 \\
\hline
\end{tabular}
\end{table}
\subsection{Kolmogorov-Smirnov test}
The best way to define the relations between the different population is to determine their luminosity function. This cannot be done in our case, since our sample are statistically incomplete. For this reason, we decided to perform a two-sample Kolmogorov-Smirnov (K-S) test to verify the compatibility between the different populations of BH masses. The null hypothesis is that the two mass distributions originate from the same population. Given two samples of $n$ and $m$ elements, the test evaluates the strongest deviation $D_n$ between their cumulative distributions and weights it by multiplying it for a corrective factor, $\sqrt{\frac{nm}{n+m}}$, which accounts for the number of sources in each sample. When the product $\mathcal{P} = D_{n}\sqrt{\frac{nm}{n+m}}$ has a lower value than a fixed threshold, the two distributions are assumed to be generated by the same population. Since we work with incomplete samples that can have a large intrinsic scatter, we decided to fix the rejection of the null hypothesis at a 99.5\% confidence level, corresponding to a threshold of $\mathcal{P} = 1.73$. To provide additional evidence for our results, we performed the test on the complete samples from the literature when possible. The results of the test are summarized in Table 8, while the cumulative distributions of our samples used in the K-S test are shown in Fig.\ref{cumulativa}. \par
\begin{table}[t!]
\label{kstest}
\caption{Two-sample Kolmogorov-Smirnov test results for BH masses. Columns: (1) First tested population; (2) number of sources $n$ in the first population; (3) second tested population; (4) number of sources $m$ in the second population; (5) $\mathcal{P}= D_n \sqrt{\frac{nm}{n+m}}$, where $D_n$ is the distance between the cumulative distributions of the populations.}
\centering
\begin{tabular}{l c l c c }
\hline\hline
Test 1  & $n$ & Test 2 & $m$ & $\mathcal{P}$ \\
\hline
F-NLS1s & 42 & Elliptical RGs &11 & 2.83 \\
F-NLS1s & 42 & S-NLS1s & 18 & 1.35 \\
F-NLS1s & 42 & RQNLS1s & 25 & 1.82 \\
F-NLS1s & 42  & Disk RGs &16 & 1.55 \\
Y08 F-NLS1s & 16 & Y08 S-NLS1s & 7 & 0.66 \\
Y08 F-NLS1s & 16 & RQNLS1s & 25 & 1.47 \\
RQNLS1s & 25 & S-NLS1s & 18 & 0.90 \\
Disk RGs & 16 & Elliptical RGs &11 & 1.81 \\
F-NLS1s &42 & Pseudobulge RGs &16 & 0.95 \\
Z06 RL & 47 & Z06 RQ & 104 & 1.63 \\
\hline
\end{tabular}
\end{table}
First of all, the K-S test between the F-NLS1s sample and the control sample of elliptical RGs reveals that their mass distributions are completely incompatible. The $\mathcal{P}$ value of 2.83 is by far higher than the fixed threshold. This expected outcome might be a sign that incomplete samples can also provide useful indications on the nature of these sources. \par
To compare flat- and steep-spectrum NLS1s, we first tested our two incomplete samples, which yielded a $\mathcal{P} =$ 1.35. This result agrees with our expectation, since it suggests that the two mass distributions are the same. To provide additional confirmation of this outcome, we tested the null hypothesis on the Y08 flux-limited sample. As mentioned in the introduction, this includes 12 flat radio spectrum and 5 steep-spectrum sources. Six other objects have unknown spectral indices. To preserve the ratio between the two groups, we therefore included four of them in the flat-spectrum and two in the steep-spectrum sample. The new result for $\mathcal{P}$ from these complete samples, 0.66, strengthens the previous result, and it allows us to conclude that F-NLS1s and S-NLS1s originate from the same population. \par
A second test was performed between the F-NLS1s and RQNLS1s. This time $\mathcal{P}$ is 1.82, which is higher than the threshold. This suggests that the two populations are intrinsically different, but when the K-S is performed between the Y08 flat-spectrum sample and the RQ sample, the result points in the opposite direction. The value of 1.47 leads us to conclude that the two distributions originate from the same population. This difference in the deviation between the complete and incomplete distributions suggests a selection effect due to the incompleteness of our F-NLS1s sample. For instance, the redshift distributions of the complete and incomplete samples are quite different, since the flat-spectrum objects are on average located farther away from us than the radio-quiet sources. Moreover, both the histogram of Fig.\ref{masse} and the cumulative distribution for F-NLS1s in Fig.\ref{cumulativa} show that F-NLS1s masses are highly concentrated between $10^{7.5-8}$ M$_\odot$, and this might be interpreted as a sign that a selection effect is also present in the flat-spectrum sample. \par
To further investigate this problem in depth, we performed a test on a larger complete sample of 2011 NLS1s derived in Zhou et al. (2006, from now on Z06). The mass values they found are not directly comparable with ours because they used a different method to derive them. For this reason, we instead cross-matched their sample with the FIRST survey \cite{Becker94}, finding all the radio-emitting sources. Then we split the resulting sample of 151 sources according to their radio-loudness, calculated as in our work. We finally compared the masses they found for the resulting samples of radio-quiet and radio-loud sources. We did not divide the radio-loud sample into steeps and flat-spectrum objects since, as shown before, they can be considered as part of the same population. The K-S test result is somewhat in the middle between our previous results, with a $\mathcal{P}$ of 1.63. This value is again below the rejection threshold, but it is higher than our result between complete samples. This might indicate that the radio-quietis actually different from the radio-loud population, but this conclusion is not so straightforward. These ambiguous results cause us to consider the K-S test as inconclusive for RQNLS1s. \par
The third test between F-NLS1s and disk RGs provides a $\mathcal{P}$ of 1.55. This is a really interesting result because it appears to relate NLS1s with sources that are usually considered a different class of AGN. At the same time the test indicates that disk RGs have a different mass distribution than elliptical RGs, since the $\mathcal{P} = 1.81$ is above the fixed threshold. Unfortunately, the disk RGs sample has no complete subsamples that we might have used to provide further confirmation of these results. 

\subsection{Resulting scenario}
\textbf{Steep-spectrum radio-loud NLS1s.} The result for incomplete samples can be considered conclusive, and the complete samples provide further confirmation with an even more definitive outcome. Both of them reveal that flat- and steep-spectrum RLNLS1s originated from the same population. The different numerical result between complete and incomplete samples is probably due to a selection effect. Figure \ref{cumulativa}, for instance, shows the distributions of the incomplete samples, and those of F-NLS1s and S-NLS1s are similar, but do not overlap systematically. The reason for this is that the flat-spectrum sample has many more sources at high $z$, while sources at low $z$ are much more common in the steep-spectrum sample. The use of the Y08 sample allowed us instead to compare sources with the same $z$ distribution. This shows that the two mass distributions become closer, confirming the high compatibility of the two samples. Another hint that flat- and steep-spectrum RLNLS1s are related is given by Fig.\ref{massacc}. The samples almost entirely overlap in the plot because they not only have a similar mass distribution, but they also show similar distributions of $\epsilon$, indicating that their accretion mechanism might be the same. \par
In conclusion, the resulting compatibility between the distributions of flat- and steep-spectrum sources in the two samples reveals that they originate from the same population. This occurs in analogy with what is observed for blazars and RGs: when a flat-spectrum source is observed under a large inclination angle $i$, the radio-lobe emission starts to dominate the emission of the core, and the radio-spectrum becomes steep. The steep-spectrum sources therefore are misaligned F-NLS1s, and as expected they are parent sources. Nevertheless, as mentioned in the introduction, there are too few of them to explain the nature of the whole parent population.\\
\textbf{Radio-quiet NLS1s.} As mentioned in the previous section, the K-S test on the radio-quiet sample cannot be considered conclusive because of its contradictory results. In particular, the result of the test between the radio-quiet sources and the F-NLS1s incomplete sample shows that they are not compatible. This result is also visible in Figs.\ref{massacc} and \ref{cumulativa}. RQNLS1s are concentrated at lower masses than the F15 sample, so the two populations appear to be distinct. The cumulative distribution of RQNLS1s is also systematically higher than that of S-NLS1s, and this is a sign that radio-quiet sources are less similar to the beamed population than to the steep-spectrum population. When the two complete samples are compared instead, the data seem to indicate a higher compatibility between these two classes of sources, even if their redshift distributions are still quite different and therefore a selection effect might still be present. The large Z06 sample of RQNLS1s and RLNLS1s reveals a slightly lower compatibility, but it does not indicate a clear separation between the samples. Therefore the result is not yet conclusive.\par
A possible explanation for this ambiguity is that all the radio-quiet samples are contaminated by sources that affect the results because they actually harbor a relativistic jet. It is true that radio-quiet sources can also exhibit jets, because radio-loudness is not an absolute parameter. As shown in Ho \& Peng (2001), its value is strongly affected by the host galaxy contribution. Many sources can move from the radio-quiet to the radio-loud domain, depending on how their optical magnitude is measured, and also depending on which corrections are applied. That some RQNLS1s show an elongated structure or other signs that seem to reveal the presence of jets, indicates that at least some of them can be part of the parent population. \par
Moreover, the mass distributions of the radio-quiet sample and of the steep-spectrum sample are very similar, with a $\mathcal{P} = 0.90$. The mathematical reason of this result is evident in Fig. \ref{cumulativa}: since their cumulative distributions are pretty close, they may originate from the same population. Therefore, if steep-spectrum radio-loud and radio-quiet sources really are the same class of objects, the latter might also be related to F-NLS1s. In conclusion at present it is neither correct to exclude the presence of jets in these sources a priori just because of their radio-quietness, nor is it correct to rule them out of the parent population. A detailed investigation is needed using new generation instruments, such as JVLA or SKA, to understand how often jets are present and if their presence can allow us to include RQNLS1s in the parent population and hence increase the low number of parent sources. A study on the radio luminosity function of the different populations (Berton et al. in prep.) will also provide useful results to solve this problem. \\
\textbf{Disk-hosted RGs.} As pointed out before, the K-S test seems to confirm the relation between F-NLS1s and disk RGs, revealing also that they are closer to the beamed sample than to elliptical RGs. As mentioned in the introduction, this can be interpreted in the frame of the unified model. If the BLR has a flattened component, the radio jet is probably perpendicular to it \cite{Giovanni09}. When the observing angle $i$ is large enough, the rotational effect of the BLR clouds can broaden the permitted lines because of the Doppler effect and cause the NLS1 to appear as a BLRG. Then, when $i$ is even larger and the line of sight intercepts the molecular torus, the nuclear regions are obscured and the source appears as an NLRG. Nevertheless, this scenario is unlikely to account for all the sources, since they show a high-mass tail that no NLS1s sample has, and this is more similar to the mass distribution of elliptical RGs. A possible solution to this problem can be seen in Fig.\ref{massacc}. Disk RGs are somewhat similar to a "bridge" connecting the low-mass and high-accretion region occupied by NLS1s to the high-mass and low-accretion where the BL Lac-like elliptical RGs are located. Some of these sources may simply be genuine Seyfert 1 or Seyfert 2 galaxies with jets viewed at large angle, while some of them might instead belong to F-NLS1s parent population and therefore be just a misaligned NLS1 with Doppler-broadened lines. In Fig.\ref{massacc} they show a very wide distribution of Eddington ratios that might be due to intrinsic differences between the objects of the sample. Some of them almost overlap with the NLS1s distribution, while others have a lower $\epsilon$, more similar to slowly accreting sources such as regular Seyferts. \par
There is another interesting possibility regarding disk RGs that involves the nature of their bulges. We do not know a priori what their bulges look like, and in particular whether they are regular bulges or pseudobulges. The majority of low-redshift NLS1s are hosted in disk galaxies with a pseudobulge \cite{Deo06,Mathur12}, therefore it is reasonable to assume that their parent population shares the same characteristic. Ho \& Kim (2014) developed a M$_{BH}-\sigma_*$ relation that can be used to calculate the BH mass in presence of a pseudobulge. Its only difference to Eq.\ref{bhsigma} is the different zero-point, which is 7.91 instead of 8.49. Therefore, under this pseudobulge hypothesis, the logarithmic masses of Table \ref{table:mass2} would be decreased by a factor 0.58, and all the mass distributions would be shifted by the same value. We tested this hypothesis by increasing the number of sources with a hypothetical pseudobulge in our disk RGs sample. The resulting $\mathcal{P}$ continues to decrease until it reaches the lowest value of 0.95 in the most extreme case of a pseudobulge in all sources. This might be a sign that a better match for the parent population of the F15 sample is not simply disk RGs, but more precisely, disk RGs with a pseudobulge. \par
In conclusion, some disk-hosted BLRGs/NLRGs, in particular those having a pseudobulge, might belong to F-NLS1s parent population. Including them in this group might help to explain the low number of parent sources. Nevertheless, their total number and the fraction of pseudobulges among them are still unknown. The host galaxy has been studied in quite a few objects, meaning that there are many sources of this type yet to be classified. Therefore a statistical study on this class of objects is needed to determine whether their population is large enough to completely fill the gap among the parent sources.\par
\section{Summary} 
We tried to unveil the nature of the parent population of flat-spectrum radio-loud NLS1s. To do this, we analyzed the optical spectra of three samples of parent population candidates, steep-spectrum radio-loud NLS1s, radio-quiet NLS1s, and disk-hosted RGs, and of a control sample of elliptical RGs. In particular we focused on the H$\beta$ and [OIII] $\lambda$5007, for type 1, type 2 and intermediate sources, to determine the BH mass and the Eddington ratio of each object. The NLS1s are all concentrated in the low-mass/high-accretion region, while elliptical radio-galaxies have systematically higher BH masses and typically lower Eddington ratios. Disk RGs instead span a wide interval of masses and Eddington ratios. \par
We performed a Kolmogorov-Smirnov test on all the samples to compare their BH mass cumulative distributions with that of the F-NLS1s population. Since our samples are statistically incomplete, these results must be taken with care, but some conclusions appear to be confirmed. In particular, the control sample has a completely different mass distribution from all the other samples. The test showed that S-NLS1s have the same mass distribution as F-NLS1s and are, as expected, the best candidates for the parent population. Disk RGs are good candidates, and even if some of them might be genuine Seyfert galaxies, data suggest that those with a low-mass and high Eddington ratio and possibly a pseudobulge might be included in the parent population. Therefore the following scenario seems to emerge: when the inclination angle $i$ increases, a beamed NLS1 appears first as a steep-spectrum NLS1. Then, with a further increase of $i$, the rotation of a flattened component in the BLR broadens the lines because of Doppler effect, and a disk-hosted BLRG appears. When finally the line of sight intercepts the molecular torus, the source turns into a type 2 AGN, and it appears as a disk-hosted NLRG. \par
Our results are inconclusive on the connection between F-NLS1s and RQNLS1s, which must also be studied at radio frequencies to determine under which conditions they can develop jets. Statistical studies on larger complete samples are also needed to understand whether the number of parent sources can reach its theoretical value when the other classes of objects are included in the parent population along with S-NLS1s.

\begin{acknowledgements}
We thank the anonymous referee for helpful comments and suggestions. This research has made use of the NASA/IPAC Extragalactic Database (NED) which is operated by the Jet Propulsion Laboratory, California Institute of Technology, under contract with the National Aeronautics and Space Administration. We acknowledge the usage of the HyperLeda database (http://leda.univ-lyon1.fr). Funding for the Sloan Digital Sky Survey has been provided by the Alfred P. Sloan Foundation, and the U.S. Department of Energy Office of Science. The SDSS web site is \texttt{http://www.sdss.org}. SDSS-III is managed by the Astrophysical Research Consortium for the Participating Institutions of the SDSS-III Collaboration including the University of Arizona, the Brazilian Participation Group, Brookhaven National Laboratory, Carnegie Mellon University, University of Florida, the French Participation Group, the German Participation Group, Harvard University, the Instituto de Astrofisica de Canarias, the Michigan State/Notre Dame/JINA Participation Group, Johns Hopkins University, Lawrence Berkeley National Laboratory, Max Planck Institute for Astrophysics, Max Planck Institute for Extraterrestrial Physics, New Mexico State University, University of Portsmouth, Princeton University, the Spanish Participation Group, University of Tokyo, University of Utah, Vanderbilt University, University of Virginia, University of Washington, and Yale University. This research has made use of the SIMBAD database, operated at CDS, Strasbourg, France.
\end{acknowledgements}

\begin{appendix}
\section{Notes on individual objects}
\small
\textit{J0010+1058} \\
This Seyfert 1 galaxy has a $\sigma$ of the H$\beta$ broad component $\sim2300$ km s$^{-1}$ and showed several flaring episodes and superluminal motion in VLBI observations that can be explained with the presence of a relativistic jet \cite{Brunthaler00,Brunthaler05}. Taylor et al. (1996) found that an exponential disk fits the NIR surface brightness of the source, while \citealp{Surace01} found a single tidal arm with high star formation extended 22 kpc to the north.\\
\textit{J0150$-$0725} \\
This is a Seyfert 2 galaxy with a possible S0 host \cite{McKernan10}. It has a strong radio emission but is unresolved on VLA scale \cite{Thean00}. It has a flat FIR-to-radio spectrum, indicative of a strong nonthermal component \cite{Heisler95}.  \\
\textit{J0316+4119}\\
This is a Seyfert 2 radio galaxy in a lenticular host \cite{Paturel03}. It was detected at very high energy, as reported in Neronov et al. (2010) and also in Kadler et al. (2012). It seems to be a low-luminosity FR I galaxy with an angle between the jet axis and the line of sight of $\theta \lesssim$ 38$^\circ$. \\
\textit{J0407+0342}\\
This is a Seyfert 2 galaxy. According to Inskip et al. (2010), the host galaxy has both a bulge and a disk component, with the first being the brighter of the two. In radio the source has a typical FR II morphology, with a weak core and bright hot spots \cite{Cohen99}. Its spectrum, retrieved in the NED archive and derived from the Low Resolution Spectrograph at TNG, has a spectral resolution of 20\AA{}, therefore the [OIII] line was not resolved.\\
\textit{J0433+0521}\\
This is a Seyfert 1 galaxy with a low H$\beta$ broad component, $\sim1500$ km s$^{-1}$. It has a confirmed FR I morphology, with jets whose total extent exceeds 760 kpc \cite{Walker87}. There is an optical jet in the same apparent direction as the radio jet \cite{Barway03}. The host galaxy was analyzed by Inskip et al. (2010), and they found that it is better reproduced with a disk+bulge model, plus a nuclear point source that contributes 33\% to the total flux.\\
\textit{J0552$-$0727} \\
This is a Seyfert 2 galaxy, hosted in an SAB0 according to the RC3 catalog. HST imaging spectroscopy of the source revealed a jet-like region of [OIII] emission extended for 1'' \cite{Mulchaey94}. It has a radio source consisting of a compact core with a flat spectrum and symmetric jets \cite{Mundell00}.\\
\textit{J0725+2957}\\
This is a Seyfert 2 galaxy with strong absorption lines in the optical spectrum. It is associated with a disk galaxy, particularly an S0, the only source in the B2 sample of this kind. The radio emission originates in two symmetric jets that form an angle of $\sim$45$^\circ$ with the Galactic disk \cite{Capetti00}. \\
\textit{J1140+1743}\\
This is a Seyfert 2 galaxy hosted in an S0 with a large-scale dust lane \cite{Noel03}. Its spectrum is extremely red, with weak emission lines. We cannot exclude that the AGN is in its last phase before reaching a quiescent state \cite{Hota12}. The source has symmetric jets, and it appears to be forming a disk. \\
\textit{J1252+5634}\\
This is an Seyfert 1.5 galaxy hosted in a spiral galaxy with large tidal arms. It is a compact steep-spectrum object with a triple structure \cite{Odea98} that shows emission-line gas aligned with the radio source \cite{Hamilton02}. The gas structure forms a double shell-like morphology, with one lobe brighter and better defined than the other \cite{Axon00}. \\
\textit{J1312+3515}\\
This is a Seyfert 1 galaxy with a H$\beta$ broad component $\sigma \sim$ 2530 km s$^{-1}$, and it is hosted in a spiral galaxy \cite{Hamilton02}. It was classified by Kellermann et al. (1989) as a flat-spectrum radio-intermediate QSO because of a relatively low radio-loudness, in agreement with our result. \\
\textit{J1324+3622}\\
This is a Seyfert 2 galaxy with weak emission lines and a red spectrum. It is hosted by a S0 galaxy with a strong nuclear dust lane. Its radio morphology is that of a FR I radio galaxy with twin jets resolved on VLA scales \cite{Noel03}.\\
\textit{J1352+3126}\\
This is a Seyfert 2 galaxy whose optical spectrum increases toward longer wavelengths. It is a postmerger object, with the merged object being consistent with a late-type spiral galaxy. The radio source is also known as 3C 293, and it exhibits a one-sided jet. The latter shows emissions in optical, NIR, and UV and has a FR II structure \cite{Floyd06}. \\
\textit{J1409$-$0302}\\
Also known as Speca, this is an AGN, possibly a Seyfert 2 galaxy, hosted by a spiral galaxy that shows signs of recent episodes of star formation. The radio source has three pairs of lobes, probably produced by an intermittent radio jet activity from the AGN \cite{Hota11}\\
\textit{J1413$-$0312}\\
Based to its optical spectrum, this source would be classified as a Seyfert 2 galaxy, with no broad component in the permitted lines and no sign of FeII. Nevertheless, on the basis of its IR spectrum, Nagar et al. (2002) classified it as a NLS1, and we included it in the S-NLS1s sample. Since its H$\beta$ line shows no sign of a broad component, we determined its physical properties using the technique for type 2 AGN. Our results show that the source has a bolometric luminosity lower than the other NLS1s, and as a consequence a lower Eddington ratio. Its BH mass is not significantly different from the others. This discrepancy can be due to the strong absorption that affects the optical spectra of this object, which might lead to an underestimation of the bolometric luminosity. When this parameter is estimated from a different spectral range, its value appears to be higher \cite{Soldi11}. \\
\textit{J1449+6316}\\
This is a Seyfert 2 galaxy. Its host has a disturbed morphology, and shows isophotal twists and two spiral arms, with a thin dust lane that crosses the nuclear region. The radio source has an FR I morphology with double-sided jets \cite{Jackson03}.\\
\textit{J1550+1120}\\
This is a Seyfert 1.5 galaxy, with a strongly asymmetric H$\beta$ profile because of a strongly redshifted broad component. The host galaxy appears to have large tidal arms and a surface brightness profile well represented with an exponential profile \cite{Hamilton02}. It also has jets, whose outer lobes show multiple bright spots \cite{Rector95}. \\
\textit{J1704+6044}\\
This is a Seyfert 1.5 galaxy with a very large H$\beta$ broad component. The host galaxy is a spiral that contains a ring that surrounds an off-center bulge \cite{Hamilton02}. The radio source is very steep and lobe-dominated; the two lobes are asymmetric, and one of them appears to be stopped by a dense environment \cite{Goodlet04}.\\
\end{appendix}
\end{document}